\documentclass[12pt,preprint]{aastex}

\def\arcsecpoint{$''\!.$}

\def\ltsim{\raisebox{-.5ex}{$\;\stackrel{<}{\sim}\;$}}
\def\gtsim{\raisebox{-.5ex}{$\;\stackrel{>}{\sim}\;$}}

\shortauthors{Crenshaw et al.}
\shorttitle{UV Spectroscopy of the Absorption in NGC~5548}

\begin{document}

\title{Simultaneous UV and X-ray Spectroscopy of the Seyfert 1 Galaxy 
NGC~5548. I. Physical Conditions in the UV Absorbers\altaffilmark{1}}

\author{D.M. Crenshaw\altaffilmark{2,3},
S.B. Kraemer\altaffilmark{4},
J.R. Gabel\altaffilmark{4},
J.S. Kaastra\altaffilmark{5},
K.C. Steenbrugge\altaffilmark{5},
A.C. Brinkman\altaffilmark{5},
J.P. Dunn \altaffilmark{2},
I.M. George\altaffilmark{6, 7},
D.A. Liedahl\altaffilmark{8},
F.B.S. Paerels\altaffilmark{9},
T.J. Turner\altaffilmark{6, 7},
and T. Yaqoob\altaffilmark{7, 10}}

\altaffiltext{1}{Based on observations made with the NASA/ESA Hubble Space 
Telescope, obtained at the Space Telescope Science Institute, which is 
operated by the Association of Universities for Research in Astronomy, Inc., 
under NASA contract NAS 5-26555. These observations are associated with 
proposal 9279.}

\altaffiltext{2}{Department of Physics and Astronomy, Georgia State 
University, Astronomy Offices, One Park Place South SE, Suite 700,
Atlanta, GA 30303}

\altaffiltext{3}{crenshaw@chara.gsu.edu}

\altaffiltext{4}{Catholic University of America and Laboratory for Astronomy
and Solar Physics, NASA's Goddard Space Flight Center, Code 681,
Greenbelt, MD  20771}

\altaffiltext{5}{SRON National Institute for Space Research Sorbonnelaan 2, 
3584 CA Utrecht, The Netherlands}

\altaffiltext{6}{Joint Center for Astrophysics, University of Maryland, 
Baltimore County, 1000 Hilltop Circle, Baltimore, MD 21250}

\altaffiltext{7}{Laboratory for High Energy Astrophysics, Code 660, NASA's 
Goddard Space Flight Center, Greenbelt, MD 20771}

\altaffiltext{8}{Physics Department, Lawrence Livermore National Laboratory, 
PO Box 808, L-41, Livermore, CA 94550}

\altaffiltext{9}{Columbia Astrophysics Laboratory, Columbia University, 538 W. 
120th Street, New York, NY 10027}

\altaffiltext{10}{Department of Physics and Astronomy, Johns Hopkins 
University, Baltimore, MD 21218}

\begin{abstract}

We present new UV spectra of the nucleus of the Seyfert 1 galaxy NGC~5548, 
which we obtained with the Space Telescope Imaging Spectrograph at high 
spectral resolution, in conjunction with simultaneous {\it Chandra X-ray 
Observatory} spectra. Taking advantage of the low UV continuum and broad 
emission-line fluxes, we have determined that the deepest UV absorption 
component covers at least a portion of the inner, high-ionization 
narrow-line region (NLR). We find nonunity covering factors in the cores of 
several kinematic components, which increase the column density measurements of 
N~V and C~IV by factors of 1.2 to 1.9 over the full-covering case; 
however, the revised 
columns have only a minor effect on the parameters derived from our 
photoionization models. For the first time, we have simultaneous N~V and C~IV 
columns for component 1 (at $-$1040 km s$^{-1}$), and find that this component 
{\it cannot} be an X-ray warm absorber, contrary to our previous claim based 
on nonsimultaneous observations. We find that models of the 
absorbers based on solar abundances severely overpredict the O~VI columns 
previously obtained with the {\it Far Ultraviolet Spectrograph}, and present 
arguments that this is not likely due to variability. However, models that 
include either enhanced nitrogen (twice solar) or dust, with strong 
depletion of carbon 
in either case, are successful in matching all of the observed ionic columns. 
These models result in substantially lower ionization parameters and total 
column densities compared to dust-free solar-abundance models, and produce 
little O~VII or O~VIII, indicating that none of the UV absorbers are X-ray warm 
absorbers.

\end{abstract}

~~~~~

\section{Introduction}

Mass outflow in the form of UV and X-ray absorbers has been shown to be a 
common phenomenon in the active nuclei of Seyfert galaxies (Reynolds 1997; 
George et al. 1998; Crenshaw et al. 1999). With the advent of the Space 
Telescope Imaging Spectrograph (STIS) on the {\it Hubble Space Telescope} 
({\it HST}), the {\it Far Ultraviolet Spectroscopic Explorer} ({\it FUSE}), 
and the {\it Chandra X-ray Observatory} ({\it CXO}), it has become possible to 
study the absorption systems at unprecedented spectral resolution. Early 
studies with these instruments (e.g., Crenshaw \& Kraemer 1999; Kaastra et al. 
2000; Kaspi et al. 2000, 2002; Kriss et al. 2000) proved successful in 
isolating 
kinematic components of the absorption and constraining the physical 
conditions in the gas. The absorbers are often variable in their ionic column 
densities, and it was quickly realized that simultaneous UV, far-UV, and 
X-ray observations provide the most valuable information, since they cover an 
enormous range in ionization, as well as address the uncertain relationship 
between the UV and X-ray absorbers. Such observations in at least two of the 
three above wave bands are available for NGC 4051 (Collinge et al. 
2001), NGC 3783 (Kaspi et al. 2002; Gabel et al. 2003), NGC 3516 (Kraemer et 
al. 2002; Netzer et al. 2002), and Mrk 509 (Kraemer et al. 2003; Yaqoob et 
al. 2003).

In Crenshaw \& Kraemer (1999, hereafter CK99), we presented the first STIS 
echelle observations of the Seyfert 1 galaxy NGC~5548. We confirmed the 
presence of five kinematic components of absorption in Ly$\alpha$, N~V, and 
C~IV at significant blueshifts with respect to the systemic velocity, which 
were first detected in Goddard High 
Resolution Spectrograph (GHRS) observations (Crenshaw et al. 1999; Mathur et 
al. 1999). Compared to the GHRS spectra, the column densities of N~V and C~IV 
in the STIS spectra had decreased substantially in components 1 and 3 (at 
respective radial velocities of $-$1040 and $-$530 km s$^{-1}$, relative to an 
adopted systemic redshift of 0.01676 [CK99]).
We attributed these variations to changes in the total column densities of 
these components, but could not rule out variable ionization for component 1.

In CK99, we calculated photoionization 
models for the UV absorption components. Based on the C~IV and N~V column 
densities 
from the STIS 1998 data, we found that the ionization parameters and total 
(ionized plus neutral) hydrogen column densities of components 2 -- 5 were too 
low to produce significant O~VII and O~VIII columns. Since we only had an 
upper limit for the C~IV column density in component 1 from the STIS 1998 
spectrum, we relied on the nonsimultaneous GHRS 
observations of C~IV and N~V to produce a model for this component. In this 
case, the column density ratio of N~V/C~IV was $\sim$18, which together with 
the actual columns yielded a high 
ionization parameter (U $=$ 2.4), as well as the observed O~VII and O~VIII 
column densities from earlier observations by {\it ASCA} (Reynolds 1997; George 
et al. 1998). On this basis, we suggested that component 1 was likely to be the 
X-ray warm absorber, with the caveat that none of the UV or X-ray 
observations were simultaneous. In our conclusions (CK99), we noted that the 
GHRS C~IV and N~V observations were obtained $\sim$6 
months apart, the {\it ASCA} observations were $\sim$3 years earlier, and 
that our findings needed to be checked with simultaneous UV and X-ray 
observations.

Based on the O~VI and H~I column densities in {\it FUSE} observations of 
NGC~5548, Brotherton et al. (2002) found that component 1 could not be an 
X-ray warm absorber, and suggested that our original claim could be due in 
fact to nonsimultaneous observations. Arav et al. (2002) examined the GHRS and 
STIS spectra, and suggested that our C~IV column densities for component 4 
were underestimated by a factor $\geq$ 4, due to covering factor effects. 
Given the importance of resolving these issues, we obtained simultaneous 
{\it HST} STIS and {\it 
Chanda} observations of NGC~5548 at significantly longer exposure times than 
those for previous observations. We present the results on the UV observations 
in this paper, and the X-ray observations in a companion paper (Steenbrugge et 
al. 2003, in preparation, hereafter Paper II).

\section{Observations}

\subsection{High-Resolution Observations}

We obtained STIS echelle spectra of the nucleus of NGC~5548 through a 
0\arcsecpoint2 x 0\arcsecpoint2 aperture on two consecutive days (2002 January 
22, 23). We list the details of these observations and previous 
high-resolution 
observations obtained by {\it HST} in Table 1. The setup of the 
new observations was identical to that of our first STIS observations of 
NGC~5548 (CK99), covering the full UV range from 1150 -- 
3119 \AA\ at a velocity resolution (FWHM) of 7 -- 10 km s$^{-1}$. Table 1 
shows we were able to devote a much larger exposure time to the 
new E140M spectra, which contain all of the previously detected intrinsic 
absorption lines. We also 
list the previous GHRS observations of the C~IV (Crenshaw et al. 1999; Mathur 
et al. 1999) and N~V (Savage, Sembach, \& Lu 1997; Crenshaw et al. 
1999) regions obtained on separate occasions; we have reanalyzed these spectra 
for this paper.

We reduced the STIS spectra using the IDL software developed at NASA's 
Goddard Space Flight Center for the STIS Instrument Definition Team.
The data reduction includes a procedure to remove the background light from 
each order using a scattered light model devised by Lindler (1998). The 
individual orders in each echelle spectrum were spliced together in the 
regions of overlap. We found no evidence for variability in the E140M spectra 
obtained on consecutive days, so we averaged these spectra together for 
further analysis.

\subsection{Low-Resolution Observations and the UV Light Curve}

To place the new spectra into context, we have retrieved all of the previous 
low-resolution spectra of NGC 5548 that contain the 1360 \AA\ continuum 
region, which are those obtained by the {\it International Ultraviolet 
Explorer} ({\it IUE}), {\it the Hopkins Ultraviolet Telescope } ({\it HUT}), 
and the Faint Object Spectrograph (FOS) on {\it HST}. We give a brief log of 
these observations in Table 2.

We retrieved the most recently processed versions of these spectra from the 
Multimission Archive at the Space Telescope Science Institute (MAST), except 
for those obtained during the {\it HST} FOS monitoring on 1993 April 
19 -- May 27, for which we used the reprocessed versions in Korista et al. 
(1995). We measured continuum fluxes by averaging the points in a bin centered 
at 1360 \AA\ (observed frame) with a width of 30 \AA, and determined the 
one-sigma flux errors from the standard deviations. For the {\it IUE} spectra, 
this method overestimates the errors (Clavel et al. 1991), and we therefore 
scaled them by a factor of 0.5 to ensure that observations taken on the same 
day agree to within the errors on average.

Figure 1 shows the UV continuum light curve of NGC 5548, which spans a 24-year 
time period. The variations are dramatic, but undersampled, except for 
the 1988/1989 {\it IUE} campaign around JD 2,447,600 (Clavel et al. 1991) and 
the 1993 {\it HST} FOS/ {\it IUE} campaign around 2,449,100 (Korista et al. 
1995). The last two observations in Figure 1 are from STIS, and demonstrate 
that the large amplitude variations in the UV have continued. Our previous 
STIS observation was 
obtained at a high continuum state [F$_{\lambda}$(1360) $=$ 6.61 ($\pm$0.31) x 
10$^{-14}$ ergs s$^{-1}$ cm$^{-2}$ \AA$^{-1}$], whereas the most recent one 
was obtained at a low state [F$_{\lambda}$(1360) $=$ 1.97 ($\pm$0.20) 
x 10$^{-14}$ ergs s$^{-1}$ cm$^{-2}$ \AA$^{-1}$]. For comparison, the all-time 
low occurred on 1992 July 5 (JD 2,448,809) at a flux level of 
F$_{\lambda}$(1360) 
$=$ 1.07 ($\pm$0.17) x 10$^{-14}$ ergs s$^{-1}$ cm$^{-2}$ \AA$^{-1}$ 
(Crenshaw, Boggess, \& Wu 1993), a factor of $\sim$2 lower than our STIS 2002 
observation.

The GHRS fluxes in Figure 1 are estimates, since the continuum regions 
surrounding the broad emission lines were not observed, due to the small 
wavelength 
coverage. In these cases, we used separate fits to the continua and broad 
emission lines from the STIS spectra, and tried different linear combinations 
of these fits until we obtained an accurate match to the observed GHRS 
spectra. 
The inferred continuum levels for the two GHRS spectra are quite different:
F$_{\lambda}$(1360) $=$ 2.7 ($\pm$0.4) x 10$^{-14}$ and  5.6 ($\pm$0.4) x 
10$^{-14}$  ergs s$^{-1}$ cm$^{-2}$ \AA$^{-1}$ for the spectra of the N~V and 
C~IV regions, respectively. This has significant ramifications, since we 
assumed in our original modeling of the intrinsic absorption (CK99) that the 
GHRS spectra were obtained in a similar state.
We will discuss the implications of this result in subsequent sections.

We have also retrieved the {\it Far-Ultraviolet Spectroscopic Explorer} 
spectrum of NGC~5548, covering the range 905 -- 1187 \AA, which was obtained 
on 2000 June 7 (JD 2,451,703) by Brotherton et al. (2002). The continuum level 
at this time 
was even lower than that of the STIS 2002 spectrum. In the region of 
overlap, the STIS 2002 spectrum has a continuum flux that is 1.5 times that 
of the {\it FUSE} spectrum. In these low continuum states, the broad 
emission-line fluxes are also low, and due to the resulting contrast, the 
contributions of the narrow-line region (NLR) to the emission-line profiles 
is much easier to determine.
In particular, the narrow components of O~VI $\lambda\lambda$1031.9, 1037.6 
were identified in the {\it FUSE} spectrum by Brotherton et al. (2002), 
and the narrow components of Ly$\alpha$ $\lambda$ 1215.7, C~IV 
$\lambda\lambda$1548.2, 1550.8, and C~III] $\lambda$1908.7 were previously 
identified and measured in the low-state FOS spectrum by Crenshaw et al. 
(1993), and are also easily detected in the STIS 2002 spectra.

\subsection{Absorption-Line Variations}

In Figure 2, we show the C~IV profile from the three high-resolution spectra 
of this region, and label the kinematic components identified in our previous 
papers (Crenshaw et al. 1999; CK99). All of the previously 
identified components are still present in the new spectrum, but the strengths 
of some components have changed. As noted in CK99, 
absorption components 1 and 3 showed evidence for a significant decrease in 
their C~IV and N~V columns between the GHRS 1996 and STIS 1998 
observations (although we could only place an 
upper limit on the C IV column for component 1 in the STIS 1998 data);
the other components were consistent with no change. This trend has continued  
$\sim$4 years later; components 1 and 3 are still highly variable, with 
substantial {\it increases} in the C IV and N~V columns since the STIS 1998 
observation (for N~V, compare Figure 3 with plots in our earlier papers).
Thus, the pattern for components 1 and 3 has been an increase in ionic 
column densities with decreasing continuum flux. However, there are only three 
observations of each ion, and this does not prove a direct connection 
between column changes and variations in the ionizing flux (see \S 4.5).

Figure 3 shows a plot of the STIS 2002 data in the regions where intrinsic 
absorption is detected. As in the past, absorption is detected in the lines of 
Ly$\alpha$ $\lambda$ 1215.7, N~V $\lambda\lambda$ 1238.8, 1242.8, and C~IV 
$\lambda\lambda$1548.2, 1550.8. At this low state, and with a higher 
signal-to-noise spectrum, we have still not detected any absorption components 
from lower ionization lines, such as Si~IV $\lambda\lambda$1393.8, 1402.8 or 
Mg~II $\lambda\lambda$ 2796.3, 3803.5.
Component 6, near the systemic velocity, is still only seen in Ly$\alpha$, and 
is likely due to the interstellar medium or 
halo of the host galaxy (CK99). Components 2 and 5 are 
relatively narrow and uncomplicated. There is considerable structure in 
components 1, 3, and 4 that has persisted over $\sim$6 years. 
Components 1 and 4 each show a strong subcomponent in their red wings, and 
component 3 appears to consist of at least three roughly equally-spaced 
subcomponents. Figure 2 shows that it is the two subcomponents 
on either side of the central trough in component 3 that have increased 
dramatically in strength in the latest STIS spectrum.

\subsection{Spectral Fits and Absorption-Line Measurements}

The procedures we used to measure the absorption lines follow those of 
Crenshaw et al. (1999). To determine the shape of the underlying emission, we 
fit cubic splines separately to the continuum and the emission lines.
For the continuum, we fit four wavelength regions unaffected by emission or 
absorption features spanning the full range of the spectrum. For this 
low-state spectrum, the broad and narrow emission components are easily 
distinguished by inflections in the profiles (see also Crenshaw et al. 1993). 
We therefore performed a single spline fit to each broad and narrow component 
of Ly$\alpha$ and C~IV, again using segments unaffected by other emission or 
absorption lines. To fit the N~V emission, which is relatively weak and blended 
with Ly$\alpha$, we reproduced the C~IV broad and narrow profiles at their 
expected N~V positions (assuming the flux ratio of N~V 
$\lambda$1238.8/$\lambda$1242.8 $=$ 2), and scaled the profiles until a 
suitable fit was obtained. We note that our philosophy is to 
use the lowest reasonable emission-line profiles needed to fit the regions 
surrounding the absorption accurately. To normalize the absorption profiles, 
we divided the observed spectra by our spline fits. As described in Crenshaw et 
al. (1999), we determined uncertainties in the absorption parameters from 
photon statistics and different reasonable placements of the continuum and 
emission-line profiles.

Figure 3 shows our fits to the spectrum. We also show the narrow 
emission-line profiles separately, in order to gauge their fluxes above zero 
as a function of
radial velocity. The narrow-line profile of C~IV (and likely N~V) has a strong 
blue tail, which is common for Seyfert galaxies (Vrtilek \& Carleton 
1985), particularly for high-ionization lines from the inner NLR (Kraemer \& 
Crenshaw 2000).
The Ly$\alpha$ profile 
shows a small bump in its red wing -- the nature of the bump is uncertain. The 
centroids of the N~V and C~IV profiles are shifted to the red, relative to 
Ly$\alpha$. This is due to the contribution from the weaker member of the 
doublets in emission, which is most noticable in N~V.

An important aspect of Figure 3 is that it shows that component 4 dips below 
the narrow emission profiles in all three ions. Arav et al. (2002) find that in 
the 
earlier STIS data, their fits to the narrow-line profiles just skim the bottom 
of the absorption troughs from component 4. These authors suggest that the 
narrow emission lines are therefore not covered by the absorption components, 
and that component 4 in particular is likely to be heavily saturated. However, 
with the advantage of having a low-state spectrum, we find that the fluxes of 
the narrow-line profiles do exceed the troughs of component 4. 
Thus, {\it at least} a portion of the NLR in this small aperture 
(0\arcsecpoint2 x 0\arcsecpoint2) is covered by absorption component 4.

Table 3 shows our measurements of the radial velocity centroids, widths 
(FWHM), and covering factors for each component. The radial velocities, 
widths, and associated one-sigma uncertainties are based on averages of the 
unblended 
components in N~V and C~IV.  A comparison with our results on the previous 
GHRS and STIS observations 
(CK99) shows that the velocity centroids have not changed 
(outside of the uncertainties) over the previous six years. However, the 
widths of components 1 and 3 have increased substantially since the first STIS 
observations: from 85 to 222 km s$^{-1}$ for component 1 and from 78 to 159 km 
s$^{-1}$ for component 3. We attribute these increases in FWHM to increases in 
the depths of 
the subcomponents, relative to the central cores. We find that 
the widths of these components at the continuum level did {\it not} change, 
which 
supports our conclusion that the increases in FWHM are due to changes in the 
ionic columns rather than changes in the coverage of velocity space by the 
components.

In addition to the above measurements, we have taken advantage of the higher 
signal-to-noise ratio of the STIS 2002 spectrum to determine the covering 
factors (C$_{los}$) for the cores of several components. We used the technique 
of Hamann et al. (1997) for doublets to determine C$_{los}$, which is the 
fraction of continuum plus emission that is occulted by the absorber in our 
line of sight. Since we 
have no information to the contrary, our measurements implicitly 
assume that the continuum, broad emission, and narrow emission are equally 
covered by each absorption component. Figure 2 shows that the only component 
of absorption in C~IV that is not blended with another is component 4. 
However, for N~V, the doublets of 2, 3, and 4 are clean (N~V$\lambda$1242.8 
component 5 is blended with N~V$\lambda$1238.8 component 1, see Crenshaw et 
al. 1999). Furthermore, the wings of these components are blended with those 
from other components. Thus, we can only reliably determine covering 
factors for the cores of components 2, 3, and 4.

Our derived covering factors are given in Table 3. Components 2, 3, and 4 
all show nonunity covering factors in their cores; previously, we had only 
been able to determine that component 4 had a nonunity covering factor 
(Crenshaw et al. 1999). The actual values in this table are from the N~V 
doublet, whereas the 
lower limits are from the residual intensities in L$\alpha$.  These values are 
consistent with our estimates or lower limits from the previous GHRS and STIS 
spectra. 
Gabel et al. (2003) show that it is possible for the total covering 
factor (i.e., that for continuum plus emission) to differ among different 
lines (e.g., C~IV and N~V). This effect arises as a result of 
different covering factors for the continuum and emission-line regions, 
combined with different contributions of the continuum and emission-line 
fluxes to the underlying emission. The only direct test that we have of this 
effect is for component 4; in this case we find that the covering factors from 
the C~IV doublet is 0.91 $\pm$ 0.03, which is the same as that from the N~V 
doublet to within the errors. For components 2 and 3, we must assume that the 
C~IV covering factors are the same as those derived for N~V. Since 
we cannot measure covering factors for components 1 and 5, we assume 
that C$_{los}$ $=$ 1 for these two components. Brotherton et al. 
(2002) independently derived covering factors for the O~VI absorption 
components, and their values agree with ours to within the uncertainties.
In addition, they obtained C$_{los}$ $=$ 0.98 ($\pm$0.12) and 0.94 ($\pm$0.14) 
for components 1 and 5, respectively, which indicates that our assumption of 
unity covering factors for these two components is a reasonable one.

To determine the ionic column densities for each component, we converted each 
normalized profile 
to optical depth as a function of radial velocity, and integrated across the 
profile, as described in Crenshaw et al. (1999). For components 2, 3 and 4, we 
determined the optical depths using the measured C$_{los}$ values and the 
formalism of Hamann et al. (1997). For the other components, we assumed 
C$_{los}$ $=$ 1. As we stated in the preceding paragraph, this appears to be a 
reasonable assumption 
for components 1 and 5, and a necessary assumption for component 6 if it is 
indeed due to the interstellar medium or halo of the host galaxy.

We list the ionic column densities for each component in Table 4. For the 
GHRS and STIS 1998 spectra, we have remeasured the columns for components 2, 3, 
and 4 assuming the covering factors we determined from the STIS 2002 spectra. 
We give upper limits to the Si~IV absorption, since they provide useful 
constraints for the photoionization models. We consider the H~I column 
densities from L$\alpha$ to be lower limits. As mentioned in CK99, the 
components are more blended in Ly$\alpha$ than those in the other lines and 
are therefore more heavily saturated. Furthermore, the H~I column densities 
from Ly$\beta$ in the FUSE spectrum (Brotherton et al. 2002) are 5 --8  times 
higher than the values in Table 1. It is unlikely that this could be due to 
variability of the H~I column alone, since we don't see evidence for strong 
variations in other lines of components 2, 4, and 5. The most straightforward 
explanation is that the Ly$\alpha$ lines are indeed heavily saturated, and our 
measurements are lower limits. It 
would be interesting to obtain a longer {\it FUSE} exposure to detect 
higher-order Lyman lines, to check for saturation in Ly$\beta$.

To investigate the effects of the nonunity covering factors on the measured 
ionic columns, we compared the values in Table 4 with those we determined 
previously for the GHRS and STIS 1998 data (CK99). This
comparison reveals that the remeasured C~IV columns are 1.2 to 1.5 times larger 
than those previously determined assuming C$_{los}$ $=$ 1. The effects on the 
N~V measurements are the same or slightly larger; the remeasured N~V columns 
are 1.2 to 1.9 times larger. 

Our earlier statements on absorption variability based on the observed spectra 
are borne out by the measurements in Table 4. Component 1 shows huge (factor 
of 6 -- 10) changes in the N~V and C~IV column densities, and component 3 
shows significant (factor of $\sim$2) variations in both ions. Other 
components show no evidence or only marginal evidence for column density 
variations, given the uncertainties.

\subsection{Discussion of Observational Results}

For the first time, we have direct measurements of the N~V and C~IV column 
densities in component 1 from simultaneous observations (the GHRS observations 
were not simultaneous, and the previous STIS observation could only place a 
lower limit on the C IV column). Thus, we are able to check our previous 
suggestion that component 1 might be the X-ray warm absorber, based on its 
high N~V/C~IV ratio ($=$ 18) from the GHRS observations. In 
Table 4, we see that component 1's N~V/C~IV ratio in the STIS 2002 spectrum is 
only 2.5, which is close to the ratios for components 2, 4, and 5. 
Based on our models (\S 4), this value does not produce significant O~VII 
or O~VIII columns, and component 1 in the STIS 2002 spectrum {\it cannot} be 
an X-ray warm absorber at that epoch. Given this finding, we can no longer 
assume
that the GHRS N~V/C~IV ratio provides an appropriate constraint on the 
physical conditions in component 1. The ionic column densities in Table 4 are 
strongly variable over 2 -- 4 years, and it is certainly possible that they 
varied over the six-month interval between the GHRS observations.
Furthermore, in \S 2.2 we showed that the GHRS C~IV and N~V observations were 
obtained at different continuum states, so that one cannot call upon a 
similar ionization state of the gas as a reason for using this ratio.
Finally, photoionization models of component 1 using simultaneous 
observations of the O~VI and Ly$\beta$ absorption in {\it 
FUSE} data (Brotherton et al. 2002) indicate that this component has an 
ionization parameter too low to be a warm absorber at the epoch of 
observation (2000 June 7). Thus, there is no longer any significant evidence 
that component 1 is, or was, a warm absorber.
Note that this does not rule out the presence of X-ray absorption from a 
component of more highly ionized gas at this radial velocity, and in fact, 
Kaastra et al. (2002) present such evidence based on C~VI 
absorption in {\it CXO} spectra at the same approximate radial velocity as 
component 1 (see \S 4.4). Furthermore, in our simultaneous HETGS spectrum, the 
high ionization Si XIV Ly$\alpha$ line is dominated by a velocity component 
compatible with UV component 1 (Paper II).

As noted earlier, Arav et al. (2002) suggest that the narrow emission lines in 
the STIS aperture are not covered by the absorption components. They conclude 
that component 4, which has the deepest troughs, is therefore much more 
saturated than it appears. In particular, they suggest that the C~IV column 
densities for component 4 are at least four times greater than we had 
previously measured 
(Crenshaw et al. 1999). However, with the new low-state STIS spectrum, we find 
that the peaks of the narrow-line profiles do extend above the troughs of 
component 4 in each line (L$\alpha$, C~IV, and N~V). Thus, {\it at least} a 
portion of the NLR is 
occulted by the UV absorber associated with component 4. As discussed earlier, 
the inclusion of nonunity covering factors does increase the measured column 
densities, but only by a factor of 1.2 to 1.9. In particular, the C~IV column 
density for component 4 increased by a factor of only 1.2 with the inclusion 
of the nonunity covering factor. Thus, we find no evidence for the severe 
underestimation of column densities suggested by Arav et al. (2002).

There are a couple of caveats that we should mention concerning the above 
discussion. First, we have implicitly assumed that the continuum source, BLR, 
and NLR are all covered by the same fraction, but it could be that only the NLR 
is partially covered. In the latter case, one could devise emission-line 
profiles for the uncovered NLR that skim the bottoms of the component 4 
troughs, and thereby produce much higher ionic column densities; however, we 
have no evidence that this is actually the case.
Second, we have not been able to accurately determine the variation of covering 
factor as a function of radial velocity for any component, due to a combination 
of blending and insufficient signal-to-noise. Thus, the column densities could 
be higher if the covering factor decreases in the wings. However, we expect 
this effect would be small, since it seems unlikely that saturation 
effects in the wings would be as large as those in the troughs, which are not 
severe. An accurate determination of the covering factor as a function of 
radial velocity should be possible with a substantially longer exposure time.

For the {\it FUSE} data, Brotherton et al. (2002) generated fits to the O VI 
emission profile assuming a ``covered NLR'' and an``uncovered NLR'' case, and 
prefer the uncovered case. However, it looks to us as though the NLR 
flux from either fit exceeds the trough of component 4, and the covered NLR is 
therefore preferable. In addition, the width 
of their narrow O VI profile is $\sim$400 km s$^{-1}$ for the uncovered case 
and $\sim$650 km s$^{-1}$ for the covered case; the latter is similar to our 
values of $\sim$700 km s$^{-1}$ for C~IV and 590 km s$^{-1}$ for Ly$\alpha$. 
Thus, it appears that the covered case is much more likely for O VI, as well 
as the other lines. However, we cannot rule out the possibility that some
fraction of the narrow O~VI emission is not covered, particularly in the large 
(30$''$ x 30$''$) {\it FUSE} aperture. 

In Figure 4, we show the {\it FUSE} spectrum in the region of the O~VI lines. 
The spectrum was normalized by dividing by the covered NLR fit of 
Brotherton et al. (2002); each member of the doublet is plotted as a function 
of radial velocity. The positions of the STIS components align well with the 
absorption troughs (at most there is a 10 km s$^{-1}$ offset). Brotherton et 
al. find that there is a velocity offset of $\sim$100 km s$^{-1}$ between the 
STIS and {\it FUSE} components, and suggest this offset is probably due to 
calibration uncertainties. The source of the discrepancy is actually the 
different systemic redshifts used; Brotherton et al. used the H~I 21-cm value 
of 0.017175, whereas we use the emission-line redshift of 0.01676 to be 
consistent with our earlier papers, which amounts to a velocity offset of 125 
km s$^{-1}$.

For components 1, 2, 4, and 5, the blue member of the O~VI doublet in Figure 4 
is clearly deeper than the red member, indicating that these lines are not 
heavily saturated. However, for component 3 (Brotherton et al.'s components 
2.5 plus 3), the two lines lie on top of each other, which suggests heavy 
saturation in O~VI across the profile and a covering factor of $\sim$0.45 (in 
agreement with Brotherton et al.'s value for component 2.5). 
We suggest that the O~VI absorption profile for component 3 is 
heavily saturated, and should therefore be considered a lower limit. The 
inferred large O~VI column is consistent with the high N~V/C~IV ratio ($=$ 9.2) 
that we find for component 3, as discussed in \S 4.2 and \S 4.3.

\section{Photoionization Models of the UV Absorbers}

Photoionization models for this study were generated using the
code CLOUDY90 (Ferland et al. 1998). We have modeled the absorbers
as matter-bounded slabs of atomic gas, irradiated
by the ionizing continuum radiation emitted by the central source.
As per convention, the models
are parameterized in terms of the ionization parameter,
U, the ratio of the density of photons with energies $\geq$ 13.6 eV
to the number density of hydrogen atoms at the illuminated face of 
the slab, and the total hydrogen column density,  
N$_{H}$ ($=$ N$_{HI}$ $+$ N$_{HII}$). 
For n$_{H}$ $\ltsim$ 
10$^{8}$ cm$^{-3}$, the predicted ionic columns densities 
are not sensitive to density. For the models, we have assumed only thermal 
broadening, since 1) the absorption line widths could be due to the
superposition of unresolved kinematic components and 2) comparison models 
assuming turbulent velocities
of $\leq$ 300 km s$^{-1}$ predict nearly identical ionic columns.
The models were deemed successful when the predicted N~V and C~IV column 
densities matched those observed to within the measurement errors.
   
For the spectral energy distribution (SED) of the ionizing
radiation, we used the X-ray continuum given in Paper II.
After correcting for Galactic reddening (E$_{B-V}$ $=$ 0.03) and 
intrinsic reddening (E$_{B-V}$ 
$=$ 0.04, see Kraemer et al. 1998), we fit the UV continuum 
fluxes from 1170 to 3000 \AA~ with a power law of index
$\alpha_{\nu}$ $\approx$ $-$1.0. We assumed that the UV continuum turns over at 
the Lyman limit, with $\alpha_{\nu}$ $=$ $-$1.4, joining the
the EUV and X-ray continua at $\sim$100 \AA~(124 eV). Figure 5 shows our 
adopted SED, which is similar to that used in CK99; tests show that 
the two SEDs yield essentially identical results.

\section{Model Results}

We use our simultaneous observations of N~V and C~IV as the principal 
constraints on the models. Additional simultaneous lower limits on H~I (from 
Ly$\alpha$) and upper limits on Si~IV are also helpful. Since the GHRS C~IV 
and N~V observations were not simultaneous and were, in fact, obtained during 
different continuum states, we do not calculate models for the GHRS data, but 
use the observed columns to investigate the absorption variability. For 
additional constraints, we make use of the simultaneous {\it CXO} 
observations, as well as the nonsimultaneous observations with 
{\it FUSE}, keeping in mind that the ionic column densities are almost 
certainly variable. For models 
of the STIS columns, we begin with the most 
well-constrained data, which are those from the 2002 observation.

\subsection{Dust-Free, Solar-Abundance Models of the STIS 2002 Data}

For the STIS 2002 data, we initially assumed that the UV absorbing gas was 
free of cosmic dust. We also assumed solar elemental abundances (see 
Grevesse \& Anders 
1989), which are, by number relative to H, as follows: He $=$ 0.1, 
C $=$ 3.4 x 10$^{-4}$, N $=$ 1.2 x 10$^{-4}$, O $=$ 6.8 x 10$^{-4}$,
Ne $=$ 1.1 x 10$^{-4}$, Mg $=$ 3.3 x 10$^{-5}$, Si $=$ 3.1 x 10$^{-5}$,
S $=$ 1.5 x 10$^{-5}$, and Fe $=$ 4.0 x 10$^{-5}$.

The resulting parameters from our best fitting models are 
listed in Table 5 (each designated as component number followed by ``solar''), 
while the predicted ionic column densities are given in 
Table 6, along with the observed values for comparison. For each component, 
we matched the observed column densities N$_{CIV}$ and N$_{NV}$ to well within 
the 
measurement errors. In each case, the model predictions for N$_{HI}$ exceeded 
the lower limits determined from Ly$\alpha$ and the predictions for 
N$_{SiIV}$ are $<<$ 10$^{12}$ cm$^{-2}$ (lower than the observational upper 
limits).

As mentioned previously, we have been able to determine simultaneous N~V and 
C~IV columns for component 1 for the first time. The derived ionization 
parameter and column density for component 1 are lower than those from CK99 by 
factors of $\sim$13 and $\sim$200, respectively, and the predicted O~VII and 
O~VIII column densities are undetectable. This 
supports our conclusion (\S 2.5) and that of Brotherton et al. (2002) 
that the gas responsible for the UV absorption in component 1 cannot produce 
the observed X-ray absorption. Again, this does not exclude the presence of 
more highly ionized gas at the same velocity.

Table 6 also lists the observed columns of H~I and O~VI from the {\it FUSE} 
spectrum (Brotherton et al. 2002) obtained $\sim$19 months earlier. Based on 
similar velocity coverages, we have added their components 0.5 and 1 together 
for our component 1, and combined their components 2.5 and 3 for our component 
3. As noted previously, we use the O~VI value for component 3 as a lower limit.
Although we were able to obtain good fits to the N~V and C~IV columns and a 
reasonable match to the H~I columns (except for component 3), the dust-free 
models predict O~VI columns far in excess of those measured, by similar 
factors of 6 -- 8. This discrepancy cannot be due to the ionizing continuum 
used in the model since, unless
there is extreme structure in the ionizing continuum in the vicinity
of the O~V (113.9 eV) and O~VI (138.1eV) threshold energies, the 
model predictions for N$_{OVI}$ do not depend strongly on SED. 
We consider it highly unlikely that variability could be the main source of the 
O~VI discrepancies, for the following reasons. 1) None of the other ions show 
significant variability in components 2, 4, and 5. 2) The {\it FUSE} spectrum 
was in a lower continuum state than that for the STIS 2002 spectrum, which 
would result in even higher predicted O~VI columns for components 1 and 3, 
according to our models. 3) It seems very unlikely that the observed O~VI 
columns would all be low by roughly the same amount as a result of extreme 
decreases in the total column density for each component during this epoch.

Thus, it seems reasonable that these 
models severely overpredict the O~VI columns, even though the 
observations are not simultaneous. Brotherton
et al. (2002) suggested that models based on N$_{NV}$/N$_{CIV}$ alone may be
unreliable due to super-solar nitrogen abundances. We 
investigate this possibility in the next section and the possibility that the 
discrepancies can be explained by depletion (primarily of carbon) into dust 
grains in \S 4.3.

\subsection{Nitrogen-Enhanced Models of the STIS 2002 Data}

We generated a second set of models with a nitrogen abundance that is two times 
solar, and assumed that the enhanced nitrogen is at the expense of carbon 
alone. This is the simplest reasonable adjustment of elemental 
abundances that can be made, for the following reason. If 
enrichment of the interstellar medium above solar abundances is due to 
intermediate-mass stars, the carbon gets converted to nitrogen first through 
the CNO process, whereas the oxygen goes to nitrogen at later times and higher 
temperatures (Maeder \& Meynet 1989). Thus, for these models, we keep the same 
abundances as before, except that C $=$ 2.2 x 10$^{-4}$ and N $=$ 2.4 x 
10$^{-4}$ by number relative to H.

The model results are listed in Tables 5 and 6 under the designation ``2xN''. 
We were able to match the observed N~V and C~IV
columns as before to well within the measurement errors (and N$_{SiIV}$ is
$<<$ 10$^{12}$ cm$^{-2}$ in all cases).
Furthermore, we have obtained a reasonable (within a factor of two) match to 
the N$_{OVI}$ for 
components 1, 2, 4, and 5, as well as a value for component 3 that is 
consistent with the lower limit. In addition, the predicted H~I column
for component 3 is much closer to the observed value from {\it FUSE}.
As can be seen in Table 5, the principal effect of enhancing nitrogen and 
depleting carbon is that the ionization parameter and column density of each 
component must be lowered substantially to match the observed N~V/C~IV ratio 
and their columns. 

The values for U and N$_{H}$ are now in approximate agreement with those 
determined by Brotherton et al. (2002), except for component 3.
Based on the high N$_{NV}$/N$_{CIV}$ ratio ($=$ 9.2) for component 3, we 
determined a much larger column 
density and ionization parameter than did Brotherton et al.
(2002), and our prediction for N$_{OVI}$ is $\sim$8 times greater than
their estimate. However, given evidence that the O~VI absorption is 
heavily saturated and the quoted column density is a lower limit, our 
prediction is consistent with the observations. Thus, for all of the 
components, the enhanced nitrogen models provide a reasonably good match to all 
of the ionic columns.

\subsection{Dusty Solar Models of the STIS 2002 Data}

We generated a third set of models that starts with solar abundances, but 
includes dust within the UV absorbers. We
assumed the following depletions from solar-abundance gas phase (see
Snow \& Witt 1996): C, 65\%; O, 25\%; Mg, 80\%; Si, 94\%; and Fe, 95\% (N is 
not depleted).  
The results are listed in Tables 5 and 6 under the designation ``dusty''. 
Again, we were able to match the observed N~V and C~IV 
columns, the N$_{OVI}$ for components 1, 2, 4, and 5, and the lower limit to 
N$_{OVI}$ in component 3. The predicted H~I columns are also close to 
the {\it FUSE} values. Thus, the dusty UV models provide an equally good match 
to the ionic columns as those provided by the enhanced nitrogen models. The 
ionization parameters are nearly identical for the two cases, whereas the N$_H$ 
columns for the dusty models are 2 -- 3 times those of the enhanced nitrogen 
models.

As noted above, in Kraemer et al. (1998) we estimated an intrinsic reddening of 
the NLR in NGC~5548 of E$_{B-V}$ $\approx$ 0.04~mag. Assuming a 
Galactic ratio of reddening to the total hydrogen column (see Shull \& van 
Steenberg 1985), this corresponds to a total hydrogen column density of 2.1 x 
10$^{20}$ cm$^{-2}$. 
This in remarkable agreement with our total model hydrogen column of 2.3 x 
10$^{20} $cm$^{-2}$. Hence, it is possible that the reddening of the continuum 
and emission lines occurs in the same gas in which the UV absorption lines 
arise.
Kraemer et al. (1998) place the high-ionization inner NLR at a distance of 
$\sim$1 pc from the central source, 
which is comfortably outside the dust sublimation radius (r$_{sub}$). Following 
Barvainis (1987), and assuming a dust
sublimation temperature of 1500~K (Salpeter 1997), r$_{sub}$ $=$ 1.3 x 
L$_{46}^{1/2}$, where
L$_{46}$ is the UV -- X-ray luminosity of the central source in units of 
10$^{46}$
erg s$^{-1}$. Based on our continuum fit, L$_{46}$ $\approx$ 0.01, and hence
r$_{sub}$ $\sim$ 0.13 pc, which is well inside the inner NLR and the 
majority of the UV absorbing gas.
  
\subsection{Constraints from X-ray Spectra}
 
From our contemporaneous {\it CXO} LETGS spectrum (see Paper II), we determined 
an O~VI column density of 2 x 10$^{16}$ cm$^{-2}$ from the absorption 
lines at 21.79 \AA\ and 22.01 \AA.
We predict a total N$_{OVI}$ $=$ 1.1 x 10$^{16}$ cm$^{-2}$ for the 
nitrogen-enhanced models and 1.8 x 10$^{16} $cm$^{-2}$ for the dusty models, 
mostly from component 3, which are close to the LETGS value. This suggests that 
at least half of the observed O~VI column comes from the absorbers detected in 
the UV.
Notably, Brotherton et al.
(2002) derive a total N$_{OVI}$ $\approx$ 4.1 x 10$^{15}$ cm$^{-2}$,
significantly below the LETGS-derived value, which we suggest is further
evidence that the O~VI lines in component 3 are saturated.
Regarding the ``classic'' warm absorbers, the total O~VII and O~VIII columns 
from the contemporaneous {\it CXO} spectrum are 5 x 10$^{17}$ cm$^{-2}$ 
and 1.5 x 10$^{18}$ cm$^{-2}$, respectively. The UV components do not 
make significant contributions to the O~VII ($<$ 15\%) and O~VIII ($<$ 1\%) 
columns, confirming the presence of more highly ionized gas than found in the 
UV.

In Kaastra et al. (2002), the C~VI Ly$\alpha$ line detected in the 1999 
LETGS spectrum was fitted using a sum
of narrow components at the velocities of the five UV absorbers.
The C~VI profile in our contemporaneous LETGS spectrum shows
less structure (see Paper II), and it was best fit with three velocity 
components at $-$1050 km s$^{-1}$, $-$520 km s$^{-1}$, and
$-$160 km s$^{-1}$. The combined C~VI column density from these components is 
1.4 x 10$^{17}$ cm$^{-2}$. Based on
our dusty UV models, only component 3 possesses a significant C~VI column
(7.4 x 10$^{15}$ cm$^{-2}$ for the nitrogen-enhanced model, 1.3 x 10$^{16}$ 
cm$^{-2}$ for the dusty model), and thus the UV components contribute $<$10\% 
of the observed C~VI.
Since C~VI requires a substantially higher ionization potential for its 
creation than O~VI (392.1 eV and 113.9 eV, respectively), this is an 
indication of more highly ionized gas {\it at the radial velocities
of the UV absorbers}, as also seen in NGC 3783 (Kaspi et al. 2002) and Mrk 509
(Yaqoob et al. 2003; Kraemer et al. 2003). 

\subsection{Changes in the Absorbers}

As noted previously, components 1 and 3 show strong variations in their ionic 
columns, whereas the other components show little evidence for 
variations. As in CK99, we can ask if the ionic column variations can be 
attributed to either variable ionization or variable total column (e.g., from
bulk motion of gas across the line of sight).
As shown in Figure 1, different UV continuum flux levels were observed
in the four high-resolution spectra of NGC 5548, with the GHRS~CIV and STIS 
1998 spectra taken during relatively 
high states and the GHRS N~V and STIS 2002 spectra taken during low states. If 
the absorbers  
respond to changes in the ionizing continuum, we would expect both 
{\it smaller} N~V and C~IV columns and {\it larger} N$_{NV}$/N$_{CIV}$ ratios
during high flux states. This is
clearly not the case for component 3; although the columns are smaller in 1998 
compared to 2002, the N~V/C~IV ratio is smaller as well (6.8 in 1998 compared 
to 9.2 in 2002). Although we only have an upper
limit for N$_{CIV}$ in the STIS 1998 data for component 1, N$_{NV}$ 
requires an ionizing continuum a factor of $\sim$6 higher during the
STIS 1998 observations. However, the observed UV flux was only 3.4 times as 
bright. Hence, either the amplitude of the variations in the EUV ionizing 
radiation is about twice as large as that in the UV, or the observed changes 
in the ionic columns for component 1 are 
not consistent with the changes in the continuum flux.
  
Given the lack of response of the absorbers to changes in the continuum flux 
(with the possible exception of component 1) we can derive upper limits to 
their electron densities and lower limits to their distances from the central 
continuum source.
The timescale for the recombination of an ion X$_{i}$
is given by the following expression (Krolik \& Kriss 1995;
Bottorff, Korista, \& Shlosman 2001):
\begin{equation}
t(X_i) = \left[\alpha(X_i)~ n_e~~ \left({f(X_{i+1})\over{f(X_i)}} - 
{\alpha(X_{i-1})\over{\alpha(X_i)}}\right)\right]^{-1},
\end{equation}
where $\alpha(X_i)$ is the recombination coefficient
for ion $X_{i}$, $f(X_i$) is the fraction of element $X$ in ionization
state $i$, and n$_{e}$ is the electron density.
Based on our model predictions, components 1, 2, 4, and 5 have similar 
conditions, with T $\approx$ 2.4 x 10$^{4}$ K, f(C~V)/f(C~IV) $\approx$ 6, and 
f(N~VI)/f(N~V) $\approx$ 1.3. For component 3, 
which is at higher ionization,
we predict  T $\approx$ 4.3 x 10$^{4}$ K, f(C~V)/f(C~IV) $\approx$ 70, 
and f(N~VI)/f(N~V) $\approx$ 16. Assuming only radiative recombination
(see Shull \& van Steenberg 1982), and a timescale of $\gtsim$ 4 yrs (the 
interval between STIS observations),
we obtain the following upper limits on electron density: for Component 3,
15 cm$^{-3}$ and 50 cm$^{-3}$ (from C~IV and N~V, respectively); for
the lower ionization components, we obtain 150 cm$^{-3}$ and 650 cm$^{-3}$.
Based on our estimate of the central source luminosity above the Lyman limit 
(see Kraemer et al. 1998) and our ionization parameters, the lower
limit to the radial distances of the low ionization components is
between 20 and 250 pc, while that for component 3 is between 150 and 250 pc
from the central source. These constraints place the absorbers at large
enough distances that they may cover much of the entire NLR. However, 
we note that the continuum and absorption variations are obviously 
not well characterized, with just a few high-resolution observations widely 
separated in time. Since the ionic columns for components 2, 4, and 5 have not 
varied significantly, despite large variations in the continuum fluxes, the 
above constraints are likely valid. For the variable components 1 and 3, 
however, we cannot rule out changes in ionization in response to 
unobserved continuum variations on time scales smaller than years. 
Thus, we consider the limits on the densities and radial locations of 
components 1 and 3 to be loose constraints; more frequent 
monitoring at high spectral resolution are needed to test them.
 
Since there is no clear signature of a response to the changes in the 
continuum flux, we can investigate whether or not the variations in ionic 
column densities are consistent with changes in the
total column densities of the absorbers. Pursuing this, we generated
sets of nitrogen-enhanced and dusty models for components 2 -- 5 in the STIS 
1998 data, with the same 
ionization parameters used for the STIS 2002 models, but adjusting the total 
column densities (see Table 7).
In each case, the predictions for N$_{CIV}$ and N$_{NV}$ match the 
measured values to within the measurement errors. Components 2, 4, and 5 show 
only small changes in the total columns, consistent with our earlier claims of 
no or marginal variability of these components. Component 3 shows a much 
smaller column in the 1998 data, and the agreement between observed and 
predicted columns indicates that the variations are likely dominated by 
changes in the total column (by a factor of $\sim$2.6). We did not attempt to 
model component 1, since we only had an upper limit for N$_{CIV}$ from the 
STIS 1998 spectra. Of course, we can reproduce the observed N~V column in 1998 
by reducing the total column of gas by a factor of $\sim$6.
Obviously, the larger C~IV and N~V columns in components 1 and 3 for the two 
GHRS observations (compared to the STIS 1998 values) could also be attributed 
to changes in the total columns, but we have no constraints on possible 
ionization changes.

\section{Summary}

New {\it HST} STIS echelle spectra, combined with simultaneous {\it CXO} 
observations, place tighter constraints on the UV absorbers of NGC~5548 than 
were possible in the past. Simultaneous detection and measurement of the C~IV 
and N~V absorption lines in component 1 show that this component is {\it 
not} an X-ray warm absorber, in agreement with the conclusion of Brotherton et 
al. (2002) based on {\it FUSE} observations of O~VI and Ly$\beta$, and 
despite our earlier claim based on nonsimultaneous GHRS observations (CK99).
The new STIS observations were obtained when the UV continuum and broad 
emission 
lines fluxes were low, allowing us to easily distinguish the narrow 
components of the emission lines. We present direct evidence that UV absorption 
component 4 covers at least a 
portion of the NLR. Our derivation of the 
covering factors indicates that our previous C~IV and N~V column 
densities were underestimated by factors of 1.2 -- 1.9, and not the $>$4
factors suggested by Arav et al. (2002).

Our dust-free, solar-abundance models of the UV absorbers in the STIS 2002 data 
match the C~IV and N~V columns well, but they severely overpredict the observed 
O~VI columns in the nonsimultaneous {\it FUSE} data. Models that include 
enhanced nitrogen or elemental depletion into dust, in both cases 
reducing the carbon in gas phase, provide reasonable (within a factor of two) 
matches to the O~VI 
columns. Absorption lines (or lack thereof) from heavier refractory elements 
that tend to be depleted into dust (e.g., Fe, Mg, Si), would be extremely 
useful for discriminating between these two models, but unfortunately there are 
no suitable high-ionization lines from these elements in the UV. In particular, 
Si~IV, with a creation ionization potential of 33.5 eV, has an insignificant 
column even in our dust-free models.

If the dusty models are correct, then the total hydrogen column from the 
absorbers is exactly the amount needed to redden the inner NLR by the observed 
amount, E(B$-$V) $=$ 0.04 (Kraemer et al. 1998), assuming a Galactic 
dust-to-gas ratio. This is a piece of evidence in favor of the dusty models, 
although it is possible that the NLR clouds provide their own reddening.
There is strong evidence for dust in UV absorbers at distances of $\gtsim$100 
pc in NGC~3227 (Crenshaw et al. 2001) and Akn~564 (Crenshaw et al. 2002).
However, in these cases, the absorbers appear at radial velocities close to 
systemic and are probably  due to the host galaxy, as opposed to the outflowing 
absorbers seen in NGC~5548.

Our absorber models indicate that none of the UV components contribute 
significantly to the O~VII and O~VIII columns seen in the X-rays, in agreement 
with the conclusions of Brotherton et al. (2002). Hence, none of the UV 
absorbers is a classic X-ray warm absorber. However, the UV components, 
particularly the higher ionization component 3, do
contribute $>$50\% of the O~VI column and 5 -- 10\% of the C~VI column 
measured in the {\it CXO} spectra (Paper II).

Regarding the source of the ionic column variations in components 1 and 3, we 
come to the same conclusions as in CK99. The variations in component 3 are 
dominated by changes in the total column density, whereas those in component 1 
are probably dominated by column changes, although we cannot rule out variable 
ionization of the gas for this component. As discussed in CK99, changes in 
total column due to bulk motion of the gas across our line of sight in NGC~5548 
require transverse velocities of $\sim$ 1500 km s$^{-1}$. 

There are a couple of indications that the UV components 2, 4, and 5 are at 
significant distances from the active nucleus of NGC~5548. We have 
presented direct evidence that component 4 covers at least a portion of the 
high-ionization, inner NLR, at a distance of $\sim$1 pc from the central 
continuum source (Kraemer et al 1998). Also, these absorbers apparently do not 
respond to continuum variations over time scales of years, which suggests that 
they are at distances between 20 and 250 pc from the nucleus.

Desirable observations in the future include coordinated UV and FUV 
observations at high spectral resolution, to check our conclusions on the 
nonsolar abundances in the gas phase of the absorbers. 
Longer exposure times on the above would be desirable, to derive covering 
factors as a function of radial velocity and investigate saturation effects in 
the wings. Monitoring observations at high spectral resolution would
place tighter constraints on the densities and distances of the absorbers.

\acknowledgments

We thank Nahum Arav for helpful discussions.
SBK and DMC acknowledge support from NASA grant NAG5-4103. 
Support for proposal 9279 was provided by NASA through a grant from the Space 
Telescope Science Institute, which is operated by the Association of 
Universities for Research in Astronomy, Inc., under NASA contract NAS 5-26555. 
Some of the data presented in this paper were obtained from the Multimission 
Archive at the Space Telescope Science Institute (MAST). Support for MAST for 
non-HST data is provided by the NASA Office of Space Science via grant 
NAG5-7584 and by other grants and contracts. The Space Research Organisation of 
the Netherlands (SRON) is supported financially by NWO, the Netherlands 
Organisation for Scientific Research.

\clearpage

\figcaption[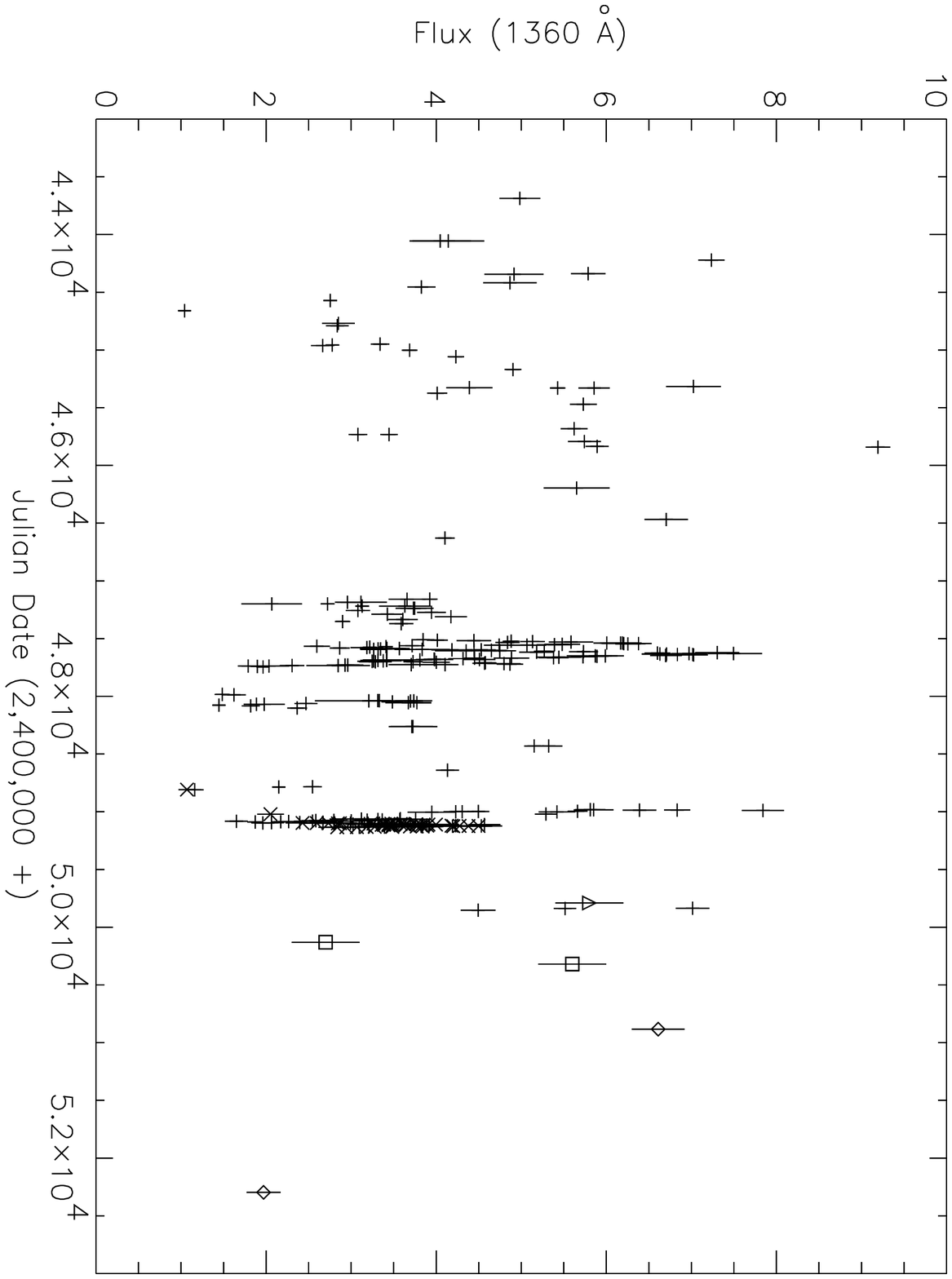]{Far-UV continuum light curve of NGC~5548. Fluxes 
(10$^{-14}$ 
ergs s$^{-1}$ cm$^{-2}$ \AA$^{-1}$) at 1360~\AA\  are plotted as a function of 
Julian date. The symbols are as follows: pluses -- {\it IUE}, X's -- FOS, 
triangle -- {\it HUT}, squares -- GHRS, diamonds - STIS. Vertical lines 
indicate the error bars ($\pm$ one sigma).}

\figcaption[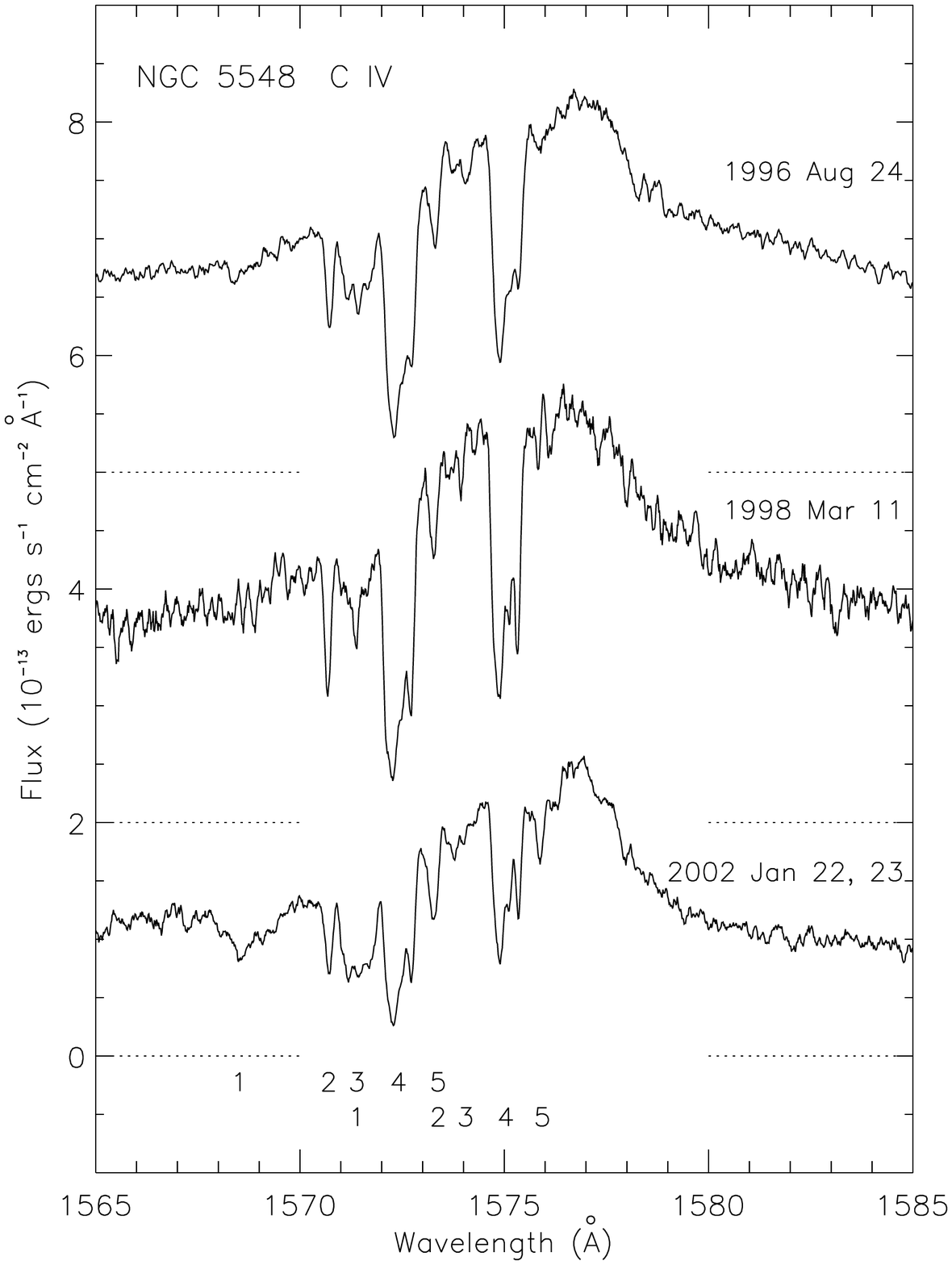]{Spectra of the C~IV region in NGC 5548 at three different 
epochs -- 1996 (GHRS), 1998 (STIS) and 2002 (STIS). The UV absorption 
components are numbered for both members of the doublet: C~IV $\lambda$ 
1548.2 and C~IV $\lambda$ 1550.8. Dotted lines give the zero flux level for 
each spectrum.}

\figcaption[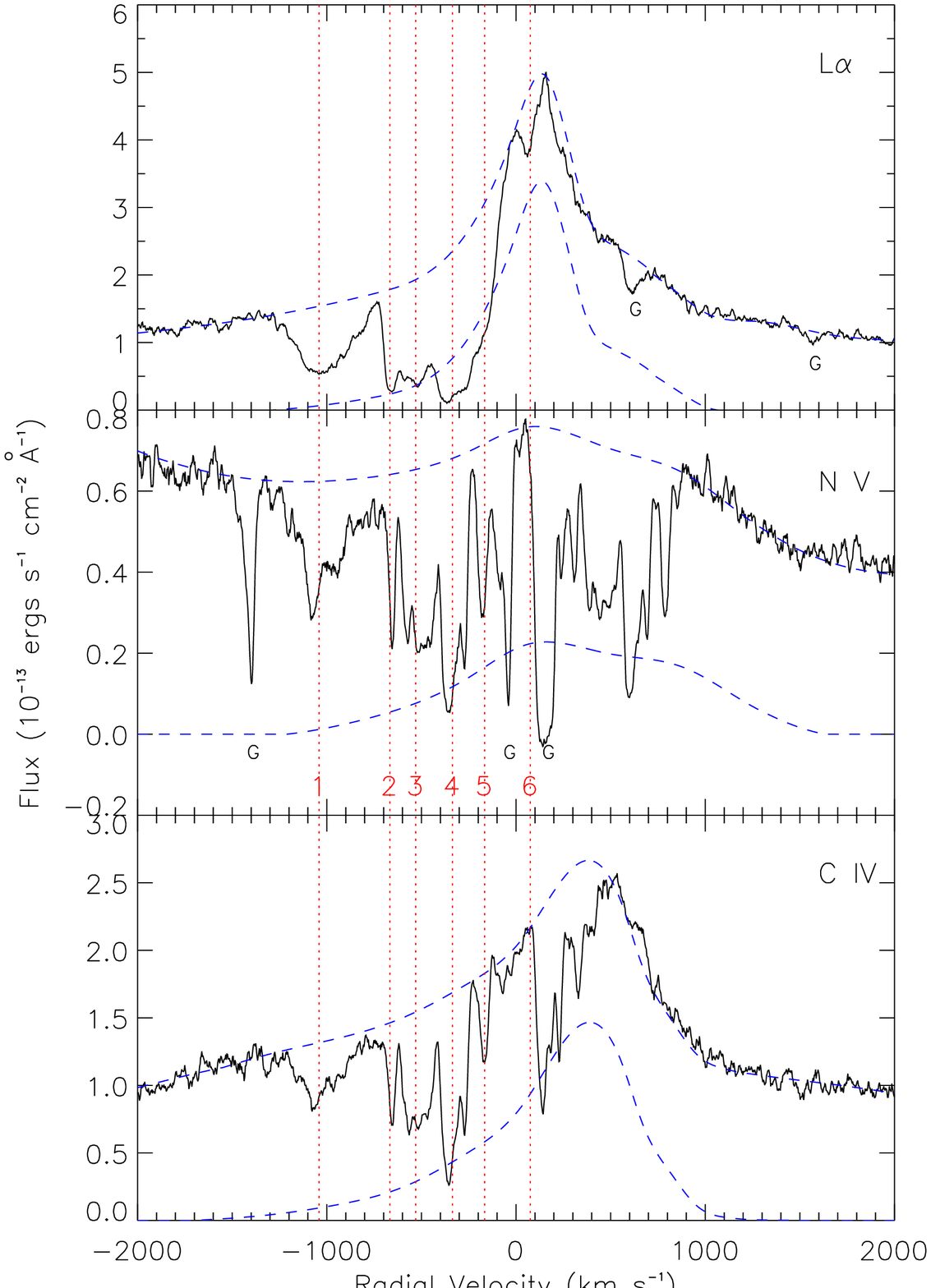]{Portions of the STIS echelle spectra of NGC~5548, showing 
the intrinsic absorption lines in different ions. Fluxes are plotted as a 
function of the radial velocity (of the strongest member, for the doublets), 
relative to an emission-line redshift of z $=$ 0.01676. The kinematic 
components are identified for the strong members of the doublets, and vertical 
dotted lines are plotted at their approximate positions. Strong Galactic 
absorption lines are labeled with ``G''. Fits to the NLR profiles are plotted 
as dashed lines.}

\figcaption[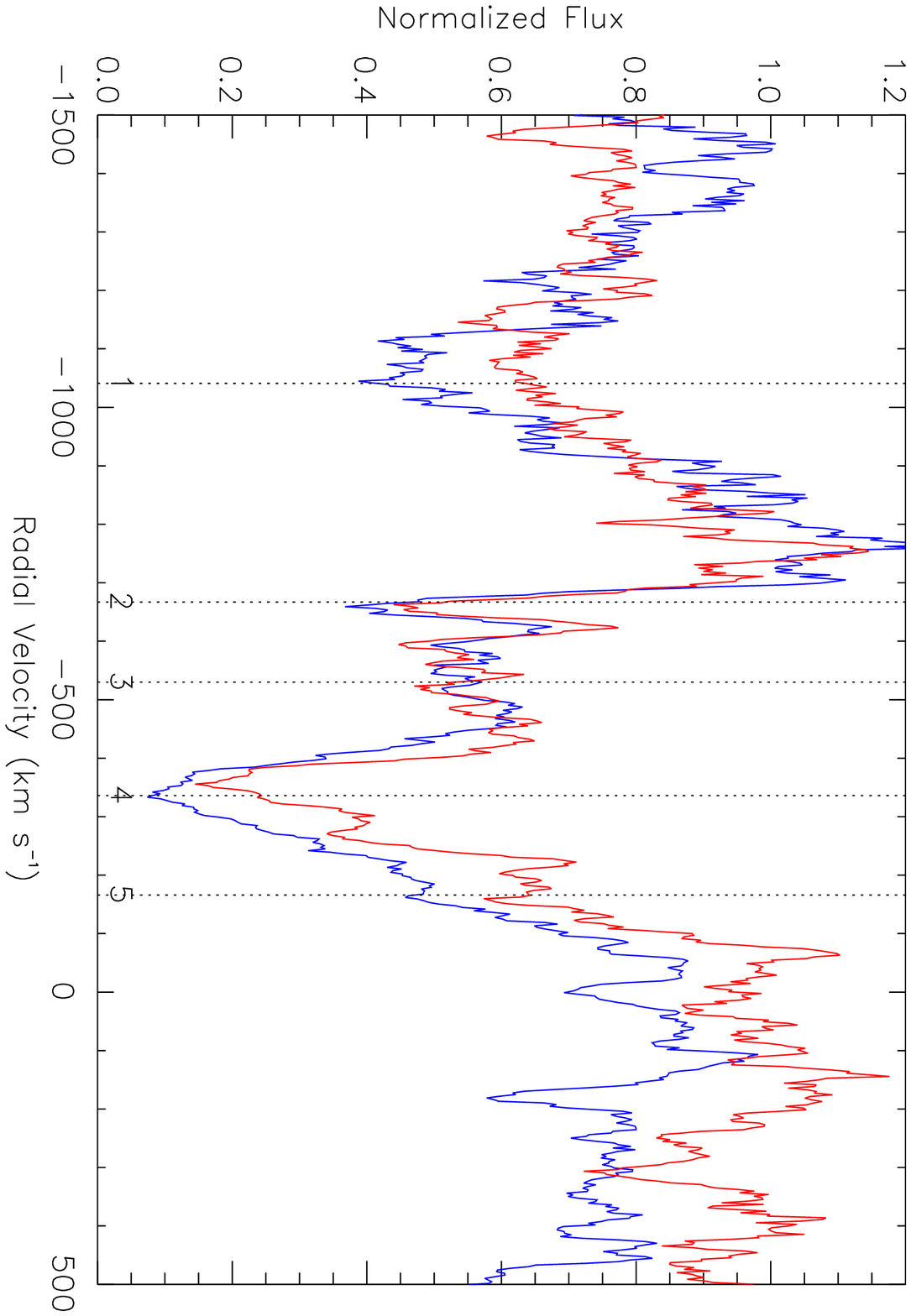]{Normalized {\it FUSE} spectra of the O~VI doublet, plotted 
as a function of the radial velocity, relative to an emission-line redshift of 
z $=$ 0.01676. The O~VI $\lambda$1031.9 components are plotted in blue, and 
the O~VI $\lambda$1037.6 components are plotted in red.}

\figcaption[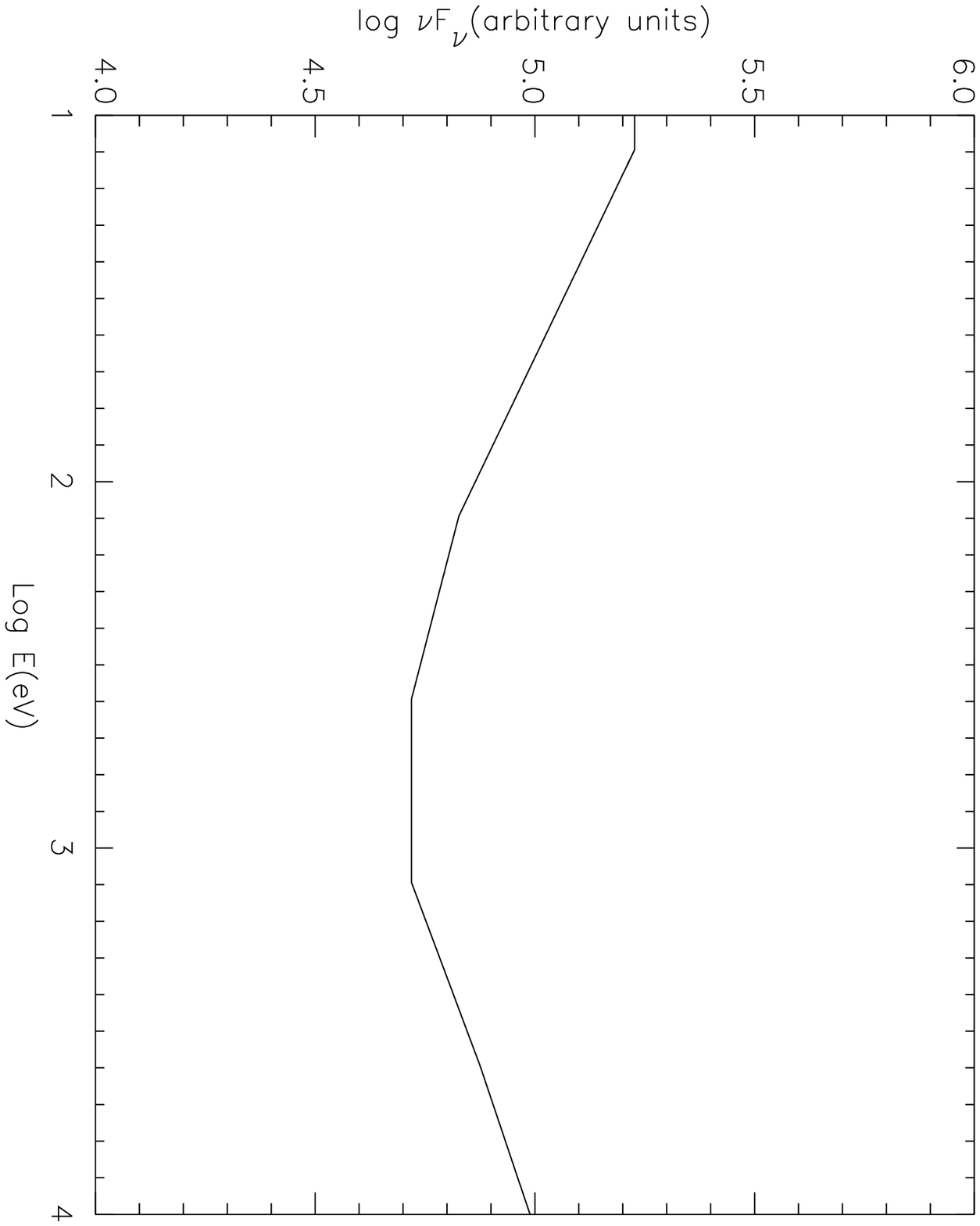]{Spectral energy distribution of the input continuum used 
for the photoionization models.}

\clearpage

\begin{deluxetable}{ccccrl}
\tablecolumns{6}
\footnotesize
\tablecaption{{\it HST} High-Resolution Spectra of NGC 5548}
\tablewidth{0pt}
\tablehead{
\colhead{Instrument} & \colhead{Grating} & \colhead{Coverage} &
\colhead{Resolution} & \colhead{Exposure} & \colhead{Date} \\
\colhead{} & \colhead{} & \colhead{(\AA)} &
\colhead{($\lambda$/$\Delta\lambda$)} & \colhead{(sec)} & \colhead{(UT)}
}
\startdata
STIS &E140M &1150 -- 1730 &46,000 &7639   &2002 January 22 \\
STIS &E230M &1607 -- 2366 &30,000 &2700   &2002 January 22 \\
STIS &E140M &1150 -- 1730 &46,000 &7639   &2002 January 23 \\
STIS &E230M &2274 -- 3119 &30,000 &2700   &2002 January 23 \\
& & & & & \\
STIS &E140M &1150 -- 1730 &46,000 &4750   &1998 March 11 \\
STIS &E230M &1607 -- 2366 &30,000 &2295   &1998 March 11 \\
STIS &E230M &2274 -- 3119 &30,000 &1905   &1998 March 11 \\
GHRS &G160M &1554$^a$ -- 1590 &20,000 &13,600 &1996 August 24 \\
GHRS &G160M &1232$^b$ -- 1269 &20,000 &4607   &1996 February 17 \\
\enddata
\end{deluxetable}
\tablenotetext{a}{Covers the C~IV region.}
\tablenotetext{b}{Covers the Ly$\alpha$ and N~V region.}

\begin{deluxetable}{lcrl}
\tablecolumns{4}
\footnotesize
\tablecaption{Low-Resolution FUV Spectra of NGC~5548}
\tablewidth{0pt}
\tablehead{
\colhead{Instrument/} & \colhead{Coverage} &
\colhead{Resolution} & \colhead{Date} \\
\colhead{Grating} & \colhead{(\AA)} &
\colhead{($\lambda$/$\Delta\lambda$)} & \colhead{(UT)}
}
\startdata
IUE SWP    &1150 -- 1978 &$\sim$250 &1978 June 27 -- 1995 May 15$^{a}$\\
FOS G130H & 1150 -- 1605 &$\sim$1200 &1992 July 5 \\
FOS G130H & 1150 -- 1605 &$\sim$1200 &1993 February 3 \\
FOS G130H & 1150 -- 1605 &$\sim$1200 &1993 April 19 -- May 27$^{b}$\\
HUT  &~820 -- 1840 &$\sim$450 &1995 March 14 \\

\tablenotetext{a}{See the IUE Merged Log at http://archive.stsci.edu/iue.}
\tablenotetext{b}{See Korista et al 1995.}
\enddata
\end{deluxetable}

\begin{deluxetable}{lccc}
\tablecolumns{4}
\footnotesize
\tablecaption{Absorption Components in NGC~5548}
\tablewidth{0pt}
\tablehead{
\colhead{Comp.} & \colhead{Velocity$^{a}$} & \colhead{FWHM} & 
\colhead{C$_{los}$$^{b}$}\\
\colhead{} &\colhead{(km s$^{-1}$)} &\colhead{(km s$^{-1}$)} & \colhead{}
}
\startdata
1        &$-$1041 ($\pm$11) & 222 ($\pm$18) & $>$0.65 \\
2        &$-$667 ($\pm$4)   & 43  ($\pm$4)  & 0.77 ($\pm$0.03) \\
3        &$-$530 ($\pm$19)  & 159 ($\pm$27) & 0.70 ($\pm$0.02) \\
4        &$-$336  ($\pm$4)  & 145 ($\pm$4)  & 0.92 ($\pm$0.02) \\
5        &$-$166 ($\pm$6)   & 61  ($\pm$13) & $>$0.69 \\
6        &$+$78 ($\pm$6  )  & 68  ($\pm$10) & $>$0.20 \\
\tablenotetext{a}{Velocity centroid for a systemic redshift of z $=$ 0.01676.}
\tablenotetext{b}{Covering factor in the line of sight.}
\enddata
\end{deluxetable}

\begin{deluxetable}{clccc}
\tablecolumns{5}
\footnotesize
\tablecaption{Measured Ionic Column Densities (10$^{14}$ cm$^{-2}$)$^{a}$}
\tablewidth{0pt}
\tablehead{
\colhead{Comp.} & \colhead{Ion} & \colhead{GHRS$^b$} & 
\colhead{STIS 1998} & \colhead{STIS 2002}
}
\startdata
1     &H~I     &---        &$>$1.34       &$>$2.31 \\
      &N V     &1.98 (0.29)   &0.44 (0.18)   &2.76 (0.10) \\
      &C IV    &0.11 (0.04)  &$<$0.17         &1.12 (0.12) \\
      &Si IV   &---       &$<$0.10        &$<$0.06 \\
& & & & \\
2     &H~I      &$>$0.76        &$>$0.61   &$>$1.18 \\
      &N V     &0.90 (0.11)   &1.11 (0.16)   &1.45 (0.13) \\
      &C IV    &0.39 (0.04)   &0.51 (0.07)   &0.56 (0.03) \\
      &Si IV   &---       &$<$0.10        &$<$0.06 \\
& & & & \\
3     &H~I      &$>$1.79        &$>$2.01   &$>$1.65 \\
      &N V     &7.40 (0.51)   &3.04 (0.16)   &8.64 (0.56) \\
      &C IV    &0.79 (0.23)   &0.45 (0.11)   &0.94 (0.26) \\
      &Si IV   &---        &$<$0.10        &$<$0.06 \\
& & & & \\
4     &H~I     &$>$3.30        &$>$2.97 &$>$3.64 \\
      &N V     &7.41 (0.87)  &10.23 (0.79)   &7.26 (0.78) \\
      &C IV    &3.43 (0.35)   &3.44 (0.47)   &2.70 (0.23) \\
      &Si IV   &---        &$<$0.10        &$<$0.06 \\
& & & & \\
5     &H~I      &$>$0.77     &$>$0.66  &$>$0.92 \\
      &N V     &1.05 (0.23)   &1.18 (0.24)    &1.45 (0.10) \\
      &C IV    &0.41 (0.18)   &0.31 (0.13)   & 0.58 (0.05) \\
      &Si IV   &---         &$<$0.10         &$<$0.06 \\
& & & & \\
6     &H~I     &0.13 (0.04)   &0.13 (0.04)    &0.13 (0.03) \\
\tablenotetext{a}{Based on C$_{los}$ values in Table 3 for components 2, 3 and 
4 and C$_{los}$ = 1 for components 1, 5, and 6. Uncertainties are given in 
parentheses. ``---'' indicates not in wavelength coverage.}
\tablenotetext{b}{GHRS C IV is not simultaneous with GHRS N V and H I.}
\enddata
\end{deluxetable}

\begin{deluxetable}{lcc}
\tablecolumns{3}
\footnotesize
\tablecaption{Photoionization Models -- STIS 2002}
\tablewidth{0pt}
\tablehead{
\colhead{Comp.} & \colhead{N$_{H}$ (cm$^{-2}$)} & \colhead{U} 
}
\startdata
1 ``solar'' & 3.3 x 10$^{19}$ & 0.18 \\
1 ``2xN''  & 3.3 x 10$^{18}$ & 0.03\\
1 ``dusty''  & 7.0 x 10$^{18}$ & 0.03\\
  & & \\
2 ``solar'' & 2.1 x 10$^{19}$ & 0.21 \\
2 ``2xN''& 1.7 x 10$^{18}$ & 0.03 \\
2 ``dusty''  & 3.8 x 10$^{18}$ & 0.03\\
  & & \\
3 ``solar'' & 1.1 x 10$^{22}$ & 1.24 \\
3 ``2xN''    & 6.6 x 10$^{19}$ & 0.22\\
3 ``dusty''  & 2.0 x 10$^{20}$ & 0.24\\
  & & \\
4 ``solar'' & 1.1 x 10$^{20}$ & 0.21  \\
4 ``2xN''    & 8.7 x 10$^{18}$ & 0.03 \\
4 ``dusty''  & 1.9 x 10$^{19}$ & 0.03 \\
  & & \\
5 ``solar'' & 1.9 x 10$^{19}$ & 0.19  \\
5 ``2xN''  & 1.9 x 10$^{18}$ & 0.03 \\
5 ``dusty''  & 3.7 x 10$^{18}$ & 0.03 \\
\enddata
\end{deluxetable}
  
\begin{deluxetable}{lccccccc}
\tablecolumns{8}
\footnotesize
\tablecaption{Predicted Ionic Column Densities (10$^{14}$ cm$^{-2}$)$^{a}$ 
--STIS 2002}
\tablewidth{0pt}
\tablehead{
\colhead{Component} & \colhead{H~I} & \colhead{N~V} & 
\colhead{C~IV} & \colhead{O~VI} & \colhead{C~VI} & \colhead{O~VII}
& \colhead{O~VIII}} 
\startdata
1 ``solar'' & 6.27 & 2.83 & 1.12 & 47.9 & 5.43 & 140 & 19.3 \\
1 ``2xN''  & 5.25    & 2.72    & 1.09   & 5.34  & 0.64   & 2.01 & 0.03 \\
1 ``dusty'' & 9.66 & 2.79 & 1.09 & 9.05 & 0.89  & 3.96 & 0.07 \\     
  &(9.41) &(2.76) &(1.12) &(7.55) & & & \\
2 ``solar'' & 3.36 & 1.45 & 0.53 & 26.7 & 35.9 & 92.6 & 15.7 \\
2 ``2xN'' & 2.74    & 1.42   &  0.57   & 2.79 &  0.37   & 1.05 &  0.01 \\
2 ``dusty'' & 5.06 & 1.50 & 0.57 & 5.03 & 0.50 & 2.28 & 0.04 \\
  &(3.25) &(1.45) &(0.56) &(3.15) & & & \\
3 ``solar''  & 100 & 8.68 & 0.98 & 486.0 & 1.0e04 & 2.1e04 & 3.6e04\\
3 ``2xN'' & 10.1   &  8.61   &  0.99  &  80.8  & 74.0  &  293.4  & 51.9  \\
3 ``dusty'' & 23.7 & 8.63 & 0.94 & 140 & 130 & 680 & 160\\
  &(16.02) &(8.64) &(0.94) &($>$10.0) & & & \\
4 ``solar'' & 16.9 & 7.39 & 2.71 & 140 & 460 & 460 & 76.7 \\
4 ``2xN'' & 13.1    & 7.21    & 2.71  & 15.2  & 1.89  &  6.13   & 0.11 \\
4 ``dusty'' & 23.9 & 7.25 & 2.70 & 25.0 & 2.53 & 11.7 & 0.24 \\
  &(20.39) &(7.26) &(2.70) &(16.92) & & & \\
5 ``solar'' & 3.39 & 1.50 & 0.57 & 26.4 & 31.4 & 130 & 12.5 \\
5 ``2xN'' &  2.83    &  1.47   &  0.58  &  2.87  & 0.35  &  1.09  &  0.02 \\
5 ``dusty'' & 5.11 & 1.47 & 0.58 & 4.78 & 0.47 & 2.10 & 0.04 \\
  &(5.60) &(1.45) &(0.58) &(3.69) & & & \\
\tablenotetext{a}{Measured values are in parentheses. The values for H~I and 
O~VI are from Brotherton et al. (2002).}
\enddata
\end{deluxetable}

\clearpage
\begin{deluxetable}{lccccc}
\tablecolumns{6}
\footnotesize
\tablecaption{Predicted Ionic Column Densities (10$^{14}$ cm$^{-2}$)$^{a}$ -- 
STIS 1998}
\tablewidth{0pt}
\tablehead{
\colhead{Component} & \colhead{U} & \colhead{N$_{H}$ (cm$^{-2}$)} & 
\colhead{H~I} & 
\colhead{N~V} & 
\colhead{C~IV}} 
\startdata
2 ``2xN''  &0.03  &1.4 x 10$^{18}$     &2.26       &1.17      &0.47 \\
2 ``dusty'' & 0.03 & 2.9 x 10$^{18}$ & 3.87 & 1.15 & 0.44 \\
& & & ($>$0.61) & (1.11) & (0.51) \\
3 ``2xN'' &0.22  &2.5 x 10$^{19}$       &3.82       &3.25      &0.37 \\
3 ``dusty'' & 0.24 & 7.8 x 10$^{19}$ & 8.61 & 3.30 & 0.35 \\
 & & & ($>$2.01) & (3.04) & (0.38) \\
4 ``2xN'' &0.03  &1.2  x 10$^{19}$    &18.32       &10.2      &3.83 \\
4 ``dusty'' & 0.03 & 2.6 x 10$^{19}$ & 33.7 & 10.2 & 3.82 \\
 & & & ($>$2.97) & (10.2) & (3.44) \\
5 ``2xN''  &0.03  &1.3 x 10$^{18}$    &2.02       &1.05      &0.41 \\
5 ``dusty'' & 0.03 & 2.8 x 10$^{18}$ & 3.87 & 1.12 & 0.44 \\
 & & & ($>$0.66) & (1.18) & (0.31) \\
\tablenotetext{a}{Measured values are in parentheses.}
\enddata
\end{deluxetable}


\clearpage
\begin{figure}
\rotatebox{90}{
\epsscale{0.8}
\plotone{f1.eps}}
\\Fig.~1.
\end{figure}

\clearpage
\begin{figure}
\epsscale{0.8}
\plotone{f2.eps}
\\Fig.~2.
\end{figure}

\clearpage
\begin{figure}
\epsscale{0.8}
\plotone{f3.eps}
\\Fig.~3.
\end{figure}

\clearpage
\begin{figure}
\rotatebox{90}{
\epsscale{0.8}
\plotone{f4.eps}}
\\Fig.~4.
\end{figure}

\clearpage
\begin{figure}
\rotatebox{90}{
\epsscale{0.8}
\plotone{f5.eps}}
\\Fig.~5.
\end{figure}
\end{document}